\documentclass[aps,prb,amsmath,amssymb,floatfix,twocolumn]{revtex4}

\usepackage{graphicx}

\begin{document}

\title{
Diamagnetism of nodal fermions 
}

\author{Amit Ghosal}
\author{Pallab Goswami}
\author{Sudip Chakravarty}
\affiliation{Department of Physics, University of California Los Angeles,
Los Angeles, CA 90095-1547}
\date{\today}

\begin{abstract}
Free nodal fermionic excitations are simple but interesting examples of  fermionic quantum criticality in which the dynamic critical exponent $z=1$, and the quasiparticles are well defined. They arise in a number of physical contexts. We derive the  scaling form of the diamagnetic susceptibility, $\chi$,  at finite temperatures and for finite chemical
potential. From  measurements in graphene, or in $\mathrm{Bi_{1-x}Sb_{x}}$ ($x=0.04$), one may be able to infer the striking  Landau diamagnetic susceptibility of the system at  the quantum critical point. Although the quasiparticles in the mean field description of the proposed $d$-density wave (DDW) condensate in high temperature superconductors is another example of nodal quasiparticles, the crossover from the high temperature behavior to the quantum critical behavior takes place at a far lower temperature due to the reduction of the velocity scale from the fermi velocity $v_{F}$ in graphene to $\sqrt{v_{F}v_{\mathrm{DDW}}}$, where $v_{\mathrm{DDW}}$ is the velocity in the direction orthogonal to the nodal direction at the Fermi point of the spectra of the DDW condensate.
\end{abstract}


\maketitle
%============================================================================
\section{Introduction}

In a class of quantum critical points (QCP), Lorentz invariance
appears as an
emergent symmetry, but in general the quasiparticle residue, as inferred from the one-particle Green's function may vanish.  In rare cases, when the quasiparticle residue is finite, depending on the statistics of the excitations, the
Lorentz invariant QCP is described by either a relativistic massless bosonic free
field theory (massless Klein Gordon action) or a relativistic massless
fermionic free field theory (massless Dirac action). Only in (1+1)-dimension
are both descriptions identical due to transmutation of statistics.
Though the theory has a relativistic form, the speed
of excitations is usually about two orders of magnitude smaller than the physical
speed of light. Due  to
fluctuations on all length scales in a critical system,
many physical quantities exhibit power laws and obey
scaling in the vicinity of the QCP. Even in the simplest of such systems, there are surprises buried in their diamagnetic response because a magnetic field is never a small perturbation: any perturbation that changes the spectra from continuous to discrete can not be considered small. Here, we hope to elaborate on this topic and present estimates that may be tested in experiments.

For  a class of  tight binding models in the half filled limit, for example graphite or graphene,  the energy vanishes
at distinct points of the Brillouin zone known as the nodal points,~\cite{Wallace,Semenoff} and in the
long wavelength and low frequency limit the dynamics are well described by
Dirac fermions obtained by linearizing the spectrum around the nodes. 
The nodal spectra can also arise from a condensate. An example is
nodal fermionic   quasiparticles of a particle-hole condensate
 in $l =2$ angular momentum channel,
as in a singlet $d$-density wave (DDW), staggered flux phase, or an orbital antiferromagnet.~\cite{DDW,Nersesyan,Nayak}

The electromagnetic charge is
a conserved quantity for a tight binding model of an electron. This is also true if the order parameter is a
particle-hole condensate, as in a DDW. In these cases, the  electromagnetic field can be
incorporated via the minimal gauge coupling. We shall restict ourselves to such systems  and not consider nodal Bogoliubov quasiparticles of a $d$-wave superconductor. The contrasting response of 
$d$-wave superconductor (DSC) and DDW is evident.~\cite{Tesanovic} The quasiparticles in a superconductor do not minimally couple to the vector
potential
$\vec A$, but to
the supercurrent
$\sim ({\vec\nabla} \varphi - 2e{\vec A}/\hbar c)$, where
$\varphi$ is the phase of the superconducting order parameter, $e$ is the electronic
charge, and $c$ is the velocity of light. 

The effect of the chemical potential, $\mu$,  is extremely important,
as it can introduce electron or hole pockets and render the linearized free
Dirac theory invalid. However, for small $\mu$ one
can still use the linearized continuum theory; $\mu=0$ describes
the vaccum of the relativistic massless theory and hence is critical. But,
for a finite $\mu$, one is dealing with a finite density
of excitations. Thus, one is perturbed away
from the criticality, and this
should provide a cutoff.

For the diamagnetic response at $\mu=0$  and zero
temperature ($T=0$), one can use a simple quantum critical scaling analysis to find the power
laws satisfied by the magnetization and the susceptibility.~\cite{Sachdev}  In this paragraph we shall set $e=\hbar=c=1$. From gauge invariance,  the vector potential $\vec{A}$ has
the same scaling dimension as the momentum which is $L^{-1}$, where $L$ is a length. Therefore  the magnetic field $H$ has the scaling dimension $L^{-2}$ or there is a length scale $L\sim H^{-1/2}$. One can immediately see that this length, which acts as a cutoff at the quantum critical point of the free Dirac fermions, is proportional to the Landau length. Since the hyperscaling should be valid for $d=2$ and the dynamic critical exponent $z=1$, the singular part of the  ground state energy density, $\Omega_{0}$, multiplied by the correlation  volume, $L^{(d+z)}$,  should be a universal number,~\cite{universal} that is, $\Omega_{0}\sim H^{3/2}$. Therefore, the magnetization behaves as $M\sim -H^{\frac{1}{2}}$ and the
diamagnetic susceptibility behaves as $\chi \sim - H^{-\frac{1}{2}}$. In the
$H\rightarrow 0$ limit, $\chi$ diverges, which  will be cut off by a number of physical effects not contained in this argument, and the stability of the state may not be in question.  

The diamagnetic sign cannot be obtained from the scaling argument. The energy levels in a magnetic field are bunched (discrete spectra) although the mean density of states is unchanged.  The number of quasiparticles  that can be accommodated below any given energy depends on whether or not this energy coincides with an eigenvalue of the Landau spectrum or falls in between two eigenvalues. For the nonrelativistic case it is easy to see that on average the energy is increased, because near $E=0$ we always start with an empty interval. For the relativistic continuum theory of nodal fermions  where we have to impose an ultraviolet cutoff, this is subtle and requires a proper regularization.  Using the work of many
authors involving $\zeta$-function  regularization,~\cite{Elizalde} we can show that the answers are indeed cutoff independent and the energy is increased. This was also checked by considering a lattice version and Peierls substitution to incorporate the magnetic field.~\cite{Nguyen} 

We cannot apply the same scaling argument to free Dirac fermions in $(3+1)$-dimensions because hyperscaling is violated. This case is best described by a mean field theory with logarithmic corrections. It is known   from explicit calculations that the singular part of the ground state energy density $\Omega_{0}\sim H^{2} \log H$.~\cite{3DDirac} A naive application of the above scaling argument gives only a regular contribution, $\Omega_{0}\sim H^{2}$, which is not surprising. Thus, we feel confident  that  the quantum critical scaling analyses are indeed meaningful.

Consider $d=2$; some aspects of the finite temperature and finite
chemical potential results can be understood from the notion of quantum criticality.   From finite size scaling, the correlation length, $\xi(T)$, is proportional to the thermal wave length, 
\begin{equation}
\lambda_{T}= \frac{\hbar v_{F}}{k_{B}T}, 
\end{equation}
that is,
\begin{equation}
\xi(T) = A_{Q}\frac{\hbar v_{F}}{k_{B}T},
\end{equation}
where $A_{Q}$ is a universal number of the order of unity and $v_{F}$ is the Fermi velocity.
Tuned to $\mu=0$, the quantum criticality will persist until $\xi(T)$ is the order of  the lattice spacing $a$.  Since $v_{F}$ is large,  one would naively expect the singular diamagnetic susceptibility $\chi \propto -H^{-1/2}$ to persist over a wide temperature range. In fact,  $\chi$ is governed by a balance between two length scales: the Landau length, $l_{B}=(\hbar c/2eB)^{1/2}$, and $\xi (T)$.
If $\xi(T)>l_{B}$, then $\chi$ follows the power law indicating the quantum critical behavior, and in the opposite limit we obtain linear response, $\chi \sim -1/T$. At $T=0$, non-zero $\mu$ tunes the system away from criticality. For small $\mu$ one can still use the linearized spectrum, and this introduces another length scale, which is essentially the interparticle spacing $\lambda\sim \hbar v_{F}/\mu$. For $\lambda>l_{B}$, $\chi$ follows a power law and in the opposite limit $\chi \sim -1/\mu$.

 Quantum criticality
of relativistic Fermions is experimentally relevant
for  graphene for which  a linear
spectrum has been established experimentally.~\cite{Novoselov} These quasiparticles
 are charged fermions and show anomalous integer quantum Hall effect,
as well as Shubnikov-de Haas oscillations.~\cite{Zhang,gusynin1,Peres} It is then natural to
expect that, as $T \rightarrow 0$, graphene should have the signature of a diamagnetic
``instability'' consistent with quantum criticality described above. Similarly,  the diamagnetic susceptibility of $\mathrm{Bi_{1-x}Sb_{x}}$ ($x=0.04$) for which the linear dispersion of the fermionic excitations is known to be present~\cite{BiSb} remains unexplored. This should be approximately describable in terms of a $(2+1)$-dimensional Dirac theory with weak interlayer coupling.~\cite{Abrikosov} As mentioned above, it has been suggested that 
the pseudogap phase of the high $T_c$ superconductors can be described by DDW, whose quasiparticle excitations  for $\mu=0$ are Dirac fermions, as was recognized a long ago.~\cite{Nersesyan} Our work is an extension of these early analyses of diamagnetism of nodal fermions to finite temperatures and finite chemical potential, which leads  to interesting results.

In a set of magnetization measurements, Ong and his collaborators~\cite{Ong} have uncovered unusual diamagnetism in the pseudogap state of the high temperature superconductor $\mathrm{Bi_{2}Sr_{2}CaCu_{2}O_{8+x}}$ (BSCCO). 
In the pseudogap regime, above the superconducting transition temperature, the diamagnetic susceptibility diverges as $\chi \sim -H^{(1-\delta)/\delta}, \; H\to 0$, where the effective exponent  $\delta(T)$ is greater than unity over a very broad range of temperature. Such a divergent susceptibility above a phase transition calls for new ideas, because the response in general should be linear. Only at a critical point, where there are fluctuations on all scales, is it possible to obtain such a nonlinearity. In particular, it is known that for two-dimensional Kosterlitz-Thouless theory $\delta=15$ at criticality,~\cite{Kosterlitz} $T= T_{KT}$, but the response is linear for any temperature  $T>T_{KT}$. To the extent that the critical region is sufficiently wide, it is of course possible to obtain a large value of susceptibility, but not   a divergent susceptibility,  as seen in measurements where fields as small as 5 Gauss were used.  Taken at its face value, experiments indicate  a critical phase extending over a wide region of  the pseudogap state. 

Long ago it was suggested that a weakly coupled stack of $XY$-systems could exhibit a floating phase in which the three-dimensional behavior at low temperatures converts to a floating power-law phase (a stack of decoupled layers) at intermediate temperatures and finally to the disordered phase at high temperatures.\cite{Tsuneto} It is now rigorously known~\cite{Toner} that if  the coupling between the layers is Josephson-like (a likely scenario), a floating phase is ruled out even for arbitrarily long-range couplings. Very special, finely tuned, interlayer couplings are necessary to produce a floating phase, which appears to be unlikely. 

Although we find that a sizable diamagnetism sets in with the DDW gap over and above the conduction electron diamagnetism, our results cannot explain the data of Ong and his coworkers: (a) there is no finite temperature critical phase; (b) the relevant scales are vastly different. As mentioned above, Kosterlitz-Thouless theory cannot account for a critical phase above $T_{c}$, though the order of magnitude is reasonably close.~\cite{Sondhi}   We hope that our calculated crossover behavior of the diamagnetic response will be  observable, at least in graphene or in $\mathrm{Bi_{1-x}Sb_{x}}$. 

The  paper is organized as follow: in Sec. II we will describe the
effective model for nodal fermions  in  two dimensions and 
outline the formalism for computing the grand thermodynamic potential. In Sec. III we will describe our results for two dimensions ($2D$). We first describe the results for the case
$\mu=0$ and then proceed to the discussion of $\mu\ne 0$. In Sec. IV we consider weak interlayer coupling in the context of a three dimensional ($3D$) system.  In Sec. V we consider numerical estimates of the effects that are experimentally relevant and in Sec. VI we conclude. There are two appendices that contain certain mathematical details.
%============================================================================
\section{Nodal Fermions: two-dimensional systems}
\subsection{Graphene}
When linearized about the two inequivalent vertices of the Brillouin zone,  the tight binding Hamiltonian, $H$, defined on a honeycomb lattice of a sheet of graphene involving only nearest neighbor hopping, with matrix element $t$, becomes in the continuum limit (lattice spacing $a\to 0$ such that $at$ is finite)

\begin{eqnarray}
H&=&\hbar v_{F}\int \frac{d^{2}k}{(2\pi)^{2}} \psi_{1}^{s\dagger}[k_{x}\sigma_{2}-k_{y}\sigma_{1}]\psi_{s1} \nonumber \\
   &  &+\hbar v_{F}\int \frac{d^{2}k}{(2\pi)^{2}} \psi_{2}^{s\dagger}[k_{x}\sigma_{2}+k_{y}\sigma_{1}]\psi_{s2},
\end{eqnarray}
where $\psi_{1}$ and $\psi_{2}$ are two species of two-component Dirac fermions corresponding to two inequivalent nodes, and $v_{F}=\sqrt{3}at/2\hbar$ is the Fermi velocity; the spin index $s$ is summed over. 
The sum over two inequivalent nodes can be written in a compact and Lorentz invariant form as
\begin{equation}
H=-i\hbar v_{F}\sum_{j=1}^{2}\int d^{2}x\bar{\psi}\gamma^{j}\partial_{j}\psi ,
\end{equation}
where $\bar{\psi}=\psi^{\dagger}\gamma^{0}$ and $\psi=\begin{pmatrix} \psi_{1} \\
\psi_{2} \end{pmatrix}$ is now a four component spinor, ignoring the irrelevant spin indices. We are using a reducible representation of $\gamma$-matrices formed from the standard Pauli-matrices $\sigma$'s:
\begin{equation}
 \gamma^{0} =\begin{pmatrix}
\sigma_3 & 0 \\ 0 &-\sigma_3
\end{pmatrix},
\gamma^{1} =\begin{pmatrix}
i \sigma_1 & 0 \\ 0 &-i\sigma_1
\end{pmatrix},
\gamma^{2} =\begin{pmatrix}
i \sigma_2 & 0 \\ 0 &-i\sigma_2
\end{pmatrix}
.
\end{equation}

The Landau level problem in the tight binding formulation is a Hofstadter problem.~\cite{Hofstadter} But, for weak enough magnetic fields we can analyze the continuum model by incorporating the magnetic field by minimal coupling prescription. So, the hamiltonian of interest takes the form
\begin{equation}
H=-i\hbar v_{F} \sum_{j=1}^{2}\int d^{2}x \bar{\psi} \gamma^{j}D_{j}\psi
\end{equation}
where $D_{j}=\partial_{j}-i\frac{e}{c}A_{j}$ is the covariant derivative.  Landau levels can be easily found by squaring the hamiltonian to be
\begin{equation}
E_{n}=\pm \frac{\hbar v_{F}}{l_{B}} \sqrt{n}\equiv\pm\sqrt{\alpha B n}
\end{equation}
where $l_{B}=(\hbar c/2eB)^{1/2}$ is the magnetic length. We have introduced
$\alpha=2\hbar ev_F^2/c$ for notational clarity. 
The same formalism can be applied
to the nodal spectra of $\mathrm{Bi_{1-x}Sb_{x}}$ ($x=0.04$).

\subsection{$d$-density wave} \label{subsec:ddw}
The nodal spectra of the DDW is also a well studied problem.~\cite{Nersesyan,Nayak}
The low-energy quasiparticle Hamiltonian for
the DDW state is 
\begin{eqnarray}
{H^{\rm DDW}} &=& \int\frac{d^{2}k}{(2\pi)^{2}}
[\left(\epsilon(k)-\mu\right)c^{s\dagger}(k)c_{s}(k) +\cr
& & {\hskip 2 cm}
i{W(k)}c^{s\dagger}(k)c_{s}(k+Q)],
\end{eqnarray}
where $\epsilon(k)$ is the single-particle energy, commonly chosen to be
\begin{eqnarray}
\label{eqn:band-structure}
\epsilon(k) = -2t(\cos{k_x}a + \cos{k_y}a)
+ 4t' \cos{k_x}a \cos{k_y}a ,
\end{eqnarray}
and $Q = (\pi/a, \pi/a)$. The nearest neighbor hopping matrix element is $t$ and the next nearest neighbor matrix element is $t'$.
The spin-singlet DDW order parameter takes the form:
\begin{equation}
\left\langle {c^{s\dagger}}({\bf k}+{\bf Q},t)\,
{c_{s'}}({\bf k},t) \right\rangle = iW(k)\,\,
{\delta^s_{s'}},
\end{equation}
where the gap function is given by
\begin{eqnarray}
\label{eqn:gap-structure}
W(k) = \frac{W_{0}(T)}{2}(\cos k_{x}a - \cos k_{y}a).
\end{eqnarray}

 As the order parameter breaks traslational invariance by a lattice spacing $a$, it is convenient 
to halve the Brillouin zone and form a two-component
Dirac spinor. Then, in the reduced Brillouin zone, the mean field Hamiltonian is
\begin{eqnarray}
H &=& \int\frac{d^{2}k}{(2\pi)^{2}}\,\chi^{s\dagger}(k)
\biggl[\frac{1}{2}(\epsilon(k) + \epsilon(k+Q)) - \mu\cr
& & \frac{1}{2}(\epsilon(k) - \epsilon(k+Q))\sigma_{3} \:+\:
W(k)\sigma_{1}\biggr]\chi_{s}(k) ,
\end{eqnarray}
where 
\begin{eqnarray}
\label{eqn:trans-indices}
\left( \begin{array}{c}
            \chi_{1s} \\ \chi_{2s}
           \end{array} \right) = \left( \begin{array}{c}
             c_{s}(k) \\ i c_{s}(k+Q)
                                         \end{array} \right)
\end{eqnarray}
The spin index $s$ can again be dropped, as this will not enter in our calculation except for an overall multiplicative factor.

The quasiparticle energies are
\begin{eqnarray}
E_{\pm}(k) &=& \frac{1}{2}(\epsilon(k) + \epsilon(k+Q))\cr
& & \pm\: \frac{1}{2}\sqrt{(\epsilon(k) - \epsilon(k+Q))^{2} + 4W^{2}(k)} .
\end{eqnarray}
At half-filling, $\mu=0$, there are 4 gapless nodal points
at $(\pm\frac{\pi}{2a}, \pm\frac{\pi}{2a})$, the
Dirac points.
A non-zero value of $\mu$ will open up fermi pockets.
The low-energy physics will be dominated by these 
gapless fermionic excitations.
We choose a single pair of nodal points,
$(\frac{\pi}{2a},\frac{\pi}{2a})$ and
$(-\frac{\pi}{2a},-\frac{\pi}{2a})$
and include the other pair
of nodes into our final result.
We take  the x-axis to be
perpendicular to the free-electron Fermi surface and the
y-axis parallel to it at one antipodal pair of nodes; similarly, the x-axis is
parallel to the free-electron Fermi surface
and the y-axis is perpendicular to it at the other pair. 
Linearizing the spectrum about the nodes,
 the dispersion relation is
\begin{equation}
E(k) = \pm \hbar \sqrt{v^{2}_{F}k^{2}_{x} + 
v^{2}_{\mathrm{DDW}}k^{2}_{y}}
\end{equation}
where ${v_F}=2\sqrt{2}\,ta/ \hbar$ and
$v_{\mathrm{DDW}}= W_{0}(T=0)a/\sqrt{2}\hbar$. It is important to note that the parameter 
$t'$ does not enter at  linear order.
It is now obvious that the formalism is identical to that described in the previous subsection  provided we  replace $v_{F}$ by $(v_{F}v_{\mathrm{DDW}})^{1/2}$ and rescale $k_{x}\to k_{x}\sqrt{v_{\mathrm{DDW}}/v_{F}}$ and $k_{y}\to k_{y}\sqrt{v_{F}/v_{\mathrm{DDW}}}$ to account
for the DDW gap anisotropy.

\subsection{Grand canonical potential}
Consider the grand canonical thermodynamic
potential per unit area of a two-dimensional (2D) system:
\begin{equation}
\Omega(T,\mu)=-k_{B}T\int_{-\infty}^{\infty} d\varepsilon D(\varepsilon)
\log\left(2\cosh \frac{\varepsilon-\mu}{2k_{B}T}\right)\cdot
\label{eq:omdef}
\end{equation}
Here, $D(\varepsilon)$ is the density of states (DOS), which in the presence
of an applied perpendicular magnetic field B, takes the following form:
\begin{equation}
D(\varepsilon)=CB\left[\delta(\varepsilon) + \sum_{n=1}^{\infty}
\left\{\delta(\varepsilon-E_n)+\delta(\varepsilon+E_n)\right\}\right]
\label{eq:dos}
\end{equation}
where $C=N_fe/h c$ is an universal constant, such that, $CB$ represents the
Landau level (LL) degeneracy factor, i.e. the magnetic flux per unit area due
to the applied field measured in the unit of flux quantum. $N_f$ is the number
of electron flavors --  $N_f=4$ for both graphene and DDW. Note that in Eq.~(\ref{eq:dos})
we have assumed a pure system. The presence of disorder broadens the sharp
$\delta$-functions in $D(\varepsilon)$, however, we restrict our discussions
to a clean system in this paper for simplicity.

Substituting $D(\varepsilon)$ in Eq.~(\ref{eq:omdef}), we can write $\Omega(\mu,T)=\Omega_0(\mu)+\Omega_T(\mu)$, where
$\Omega_0(\mu)$ is the temperature independent part (hence contributes
even at $T=0$) given by
\begin{eqnarray}
\frac{\Omega_0(\mu)}{CB}&=&\sum_{n=n_{c}+1}^{\infty}(\mu-E_n)\nonumber \\
&=&-\mu(n_c+\frac{1}{2})-\sqrt{\alpha}B^{1/2}\zeta(-\frac{1}{2},1+n_c)\cdot
\label{eq:Ommu0}
\end{eqnarray}
Here we assumed $\mu>0$ (electron doping), and thus the positive LL's are
filled only up to $n_c={\mathrm{Int}}[\mu^2/\alpha B]$ at $T=0$ while all the
negative LL's are filled (${\mathrm{Int}}[\cdot]$ stands for the `integer part'). Here 
$\zeta(s,q)=\sum_{k=0}^{\infty}(k+q)^{-s}$ is the standard Hurwitz $\zeta$-function.
It is straightforward to  $\mu<0$.
The $T$-dependent contribution is
\begin{equation}
\begin{split}
&\frac{\Omega_T(\mu)}{CBk_{B}T}=-\bigg[\log(1+e^{-\frac{\mu}{k_{B}T}})+\sum_{n=1}^{\infty} \log(1+e^{-\frac{E_n+\mu}{k_{B}T}})\\
&\quad +\sum_{n=1}^{n_c} \log(1+e^{-\frac{\mu-E_n}{k_{B}T}}) +\sum_{n=n_c+1}^{\infty} \log(1+e^{-\frac{E_n-\mu}{k_{B}T}})\bigg] .
\label{eq:OmmuT}
\end{split}
\end{equation}
Note that at finite $T$ the thermal energy can excite electrons across
$\mu$ to arbitrarily high (positive) LLs, and thus the $n$-sum
must include the whole of Dirac cone, as shown explicitly in
Eq.~(\ref{eq:OmmuT}).
%===========================================================================
\section{Results: Two dimensions}\label{sec:result}
\subsection{Undoped system, $\mu=0$}

Conside the half-filled system: $\mu=0$, hence $n_c=0$.\cite{Nersesyan} At any temperature, the length
scale of the critical fluctuations is the correlation length
$\xi(T)$. Thus, in order to observe the $T=0$ critical behavior
the largest length scale for the system must be this length. In the presence
of a magnetic field $B$, the response of the system will show critical
behavior only when $\xi(T) > l_{B}$. At $T=0$, this condition is trivially
satisfied, because the length scale of the critical fluctuations is infinite,
and we obtain 
\begin{eqnarray}
\Omega_0&=&-C\sqrt{\alpha}B^{3/2} \sum_{n=1}^{\infty} \sqrt{n} \\
&=&-C\alpha B^{3/2}\zeta(-\frac{1}{2})\\
&=&\frac{C\sqrt{\alpha}B^{3/2}\zeta(3/2)}{4\pi}\\
&=&\frac{4}{3}{\cal N}_{0}N_{f}\mathrm
{g_{2D}}\mu_{B}^{2}\sqrt{B_0}B^{3/2} .
\label{eq:om0}
\end{eqnarray}
Here $\zeta(s)=\sum_{k=1}^{\infty}k^{-s}$ is the Riemann $\zeta$-function,
${\cal N}_{0}=3\zeta(3/2)/8\pi\approx0.312$, $\mu_B=e\hbar/2mc$, the Bohr magneton, with $m$ the free electron mass. The scale $B_{0}=mv_{F}^{2}/\mu_{B}$ is a material dependent constant and has the dimension of a magnetic field. The transition from the second expression to the third is an example of   standard $\zeta$-function regularization of a divergent sum over $n$. The proof follows from the remarkable result due to Riemann,~\cite{Whittaker} that
\begin{equation}
2^{1-s}\Gamma(s)\zeta(s)\cos \left(\frac{1}{2}s\pi\right)=\pi^{s}\zeta(1-s)
\end{equation}
The logic is that the ``divergent'' sum is physically cut off at some value of $n$ and is not truly divergent, but a gauge invariant  regularization is necessary. This is accomplished by the analytic continuation given by the Riemann reflection in principle. Other regularizations are given in Refs.~\onlinecite{Nersesyan} and~\onlinecite{Nguyen}.

For reasons of physical transparency we shall often express our formulas in terms of an {\em equivalent non-relativistic free electron gas}, while keeping in mind that the real parameters that enter our calculations, such as $v_{F}$, $N_{f}$, etc. bear no real relation  to this free electron system with a circular Fermi surface and two flavors of spin. Thus, we have written
\begin{equation}
g_{\mathrm{2D}}=m/\pi\hbar^2,
\end{equation}
which is the standard, energy independent DOS of a two-dimensional ($2D$) non-relativistic Fermi gas.  Similarly, we can express
\begin{equation}
\rho_{\mathrm{2D}}=g_{\mathrm{2D}}mv_{F}^{2}/2
\end{equation}
where $\mathrm{\rho_{2D}}$ is the $2D$ arial density. Here we have used the transcription $v_{F}=\hbar k_{F}/m$, where i $k_{F}$ is the Fermi wave vector of the equivalent non-relativistic Fermi gas.

The corresponding $T\neq 0$ contribution takes the following form:
\begin{equation}
\Omega_T=-\frac{k_{B}T}{l_{B}^{2}}\left[\log2 + \sum_{n=1}^{\infty}\log\left(1+e^{-\lambda_{T}\sqrt{n}/l_{B}}
\right)\right]\cdot
\label{eq:omT}
\end{equation}
It becomes clear from Eq.~(\ref{eq:omT}) that $\Omega_T$ is a
function of the ratio of the two fundamental length scales $\lambda_{T}/l_{B}$,
and thus it must have a scaling form.

We  calculate the magnetization $M$ and the susceptibility $\chi$ from
\begin{eqnarray}
M&=&-\partial \Omega/\partial B, \\
\chi&=&\partial M/\partial H,
\end{eqnarray}
where $H$ is the magnetic field strength. These also have scaling forms. If we introduce
\begin{equation}
b=\lambda_T/l_B,
\end{equation}
 we obtain:
\begin{eqnarray}
\chi &=& \chi_0+\chi_T \\
\label{eq:netchi}
\chi_0 &=& -\frac{3C\sqrt{\alpha}}{4\sqrt{B}} \frac{\zeta(3/2)}{4\pi}=-{\cal N}_{0}N_{F}\mathrm
{g_{2D}}\mu_{B}^{2}\left(\frac{B_{0}}{B}\right)^{1/2} \\
\label{eq:chisum0}
\chi_T&=&\chi_{0}\frac{4\pi}{\zeta(3/2)}\{S+\frac{b}{3}(\partial S/\partial b)\}=\chi_{0}f(b),
\label{eq:chisumT}
\end{eqnarray}
where $S$ is given by
\begin{equation}
S=2\sum_{n=1}^{\infty} \frac{\sqrt{n}}{1+e^{b\sqrt{n}}}\cdot
\label{eq:Ssum}
\end{equation}
The function $f(b)$ defined in Eq.~(\ref{eq:chisumT}) is a universal
function of its dimensionless argument, which can be written as a series
expansion in $b$ (See Appendix A):
\begin{eqnarray}
f(b)&=&-1+\frac{4\pi}{\zeta(3/2)}\bigg[\frac{b}{18}-\frac{8}{3}\sum_{q=1}^{\infty}
\frac{b^{4q+1}}{(4q+1)!} \nonumber\\
&\times&(q+1)\eta(-4q-1)\zeta(-2q-1)\bigg]
\label{eq:fsum}
\end{eqnarray}
where, $\eta(s)=[1-2^{1-s}]\zeta(s)$ is standard Dirichlet $\eta$-function.

In the limit $\lambda_T \gg l_B$, or equivalently $b \gg 1$, it is the
quantum criticality that dictates the response of the system, and
Eq.~(\ref{eq:fsum}) is not particularly useful. Instead, we can obtain the
analytic expression for $f(b)$ in this regime by replacing
$S=2\sum_n \sqrt{n} \exp(-b\sqrt{n})$ in Eq.~(\ref{eq:Ssum}) to get (See
Appendix B for details)
\begin{equation}
f(b)=F_{3/2}(b)-b^2 F_{5/2}(b)+\frac{b^4}{12} F_{7/2}(b),
\label{eq:Fblgb}
\end{equation}
where we have defined a (convergent) $b$-dependent integral $F_p(b)$ as,
\begin{equation}
F_p(b)=\frac{4\sqrt{\pi}}{\zeta(3/2)} \int_0^{\infty} dx \frac{e^{-b^2/4x} x^{-p}}{e^x-1}\cdot
\label{eq:Fdef}
\end{equation}
In fact, it is possible to obtain an explicit $b$-dependence of $\chi$
by estimating the saddle-point approximation of $F_p$'s, which
results in
\begin{eqnarray}
&& \chi(\lambda_T \gg l_B) \approx \chi_0\left[1+\frac{1}{b}
\left\{\frac{N_{3/2}}{(e^{b^2/6}-1)}\right.\right. \nonumber\\
&&\left.\left. -\frac{N_{5/2}}{(e^{b^2/10}-1)}+\frac{N_{7/2}}{12(e^{b^2/14}-1)}\right\} \right]
\label{eq:XTlgb}
\end{eqnarray}
where, $N_p$ is a pure constant given by
\begin{equation}
N_p=\frac{\pi}{\zeta(3/2)}\sqrt{\frac{1}{2p^3}}\left\{1+{\rm erf}(\sqrt{p/2})\right\}
e^{-p[1-\log(4p)]}
\label{eq:Ndef}
\end{equation}
The message from Eq.~(\ref{eq:XTlgb})  is transparent: for $b \gg 1$ the first
term, $\chi_0$, dominates, causing the $B^{-1/2}$ behavior in the
susceptibility, while the rest of the terms in $\chi_T$ vanish exponentially.
As $b$ is decreased, $\chi_T$ grows, modifying the
non-linearity of $\chi$ in $B$. This behavior continues until $b \sim 1$, that is, until
$\xi(T)\sim\lambda_T \sim l_B$. Finally, for $b \leq 1$ the critical fluctuations fail
to describe the magnetic response, and the susceptibility follows linear response.
 For $b \ll 1$, can we keep only the first two terms in Eq.~(\ref{eq:fsum});
the next term is $\sim b^5$ and hence negligibly small. The first term
exactly cancels $\chi_0$ and we have,
\begin{equation}
\chi=-\frac{C\alpha}{24k_{B}T}=-N_{f}\frac{\mathrm{\rho_{2D}}\mu_B^2}{3k_{B}T},
\label{eq:Xsmallb}
\end{equation}
It is the expected diamagnetic, $B$-independent
behavior in the high temperature limit if we absorb $N_{f}$ in the definition of the areal density.\cite{Huang} 
\begin{figure}[htb]
\includegraphics[scale=0.425]{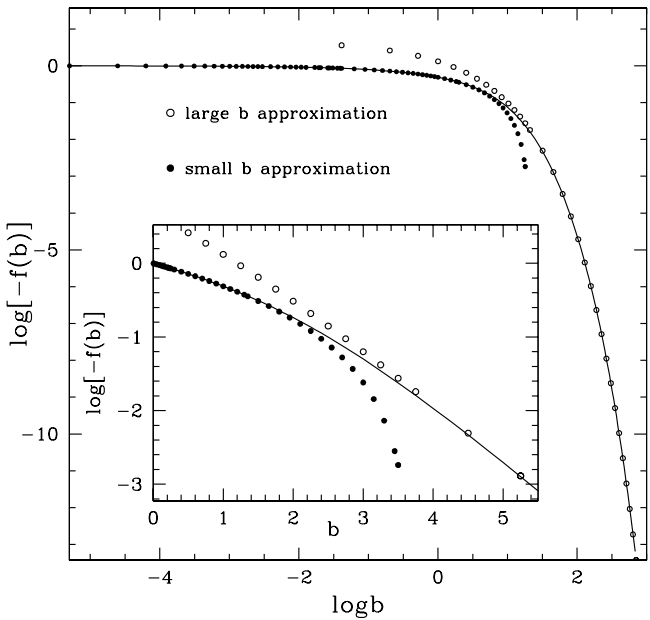}
\caption{
$\log[-f(b)]$ as a function of $\log b$. The numerical evaluation
of Eq~(\ref{eq:Ssum}) is given by solid line, and the analytical expression
for the large and small $b$ limit are given by empty and solid circles
respectively. The inset shows the blown-up crossover region (in linear
scale). The two asymptotic limits reproduce $f(b)$ surprisingly well over
almost the entire parameter regime.}
\label{fig:Fig1}
\end{figure}

We plot $\log f(b)$ as a function of $\log b$ in Fig. 1, using three following methods:
(a) by numerically evaluating Eq.~(\ref{eq:Ssum}) with a desired (high)
accuracy for a wide range of $b$;
(b) from the large $b$ asymptotic expression as in Eq.~(\ref{eq:XTlgb}), and
(c) evaluating Eq.~(\ref{eq:fsum}) in the limit $b \ll 1$, which amounts to
keeping only terms up to linear order  in $b$. We find that the two asymptotic
expressions encompass almost the entire parameter space surprisingly well.
The smoothness of $f(b)$ implies that while $\chi \sim B^{-1/2}$ for $b \gg 1$,
its behavior smoothly crosses over to $B$-independent diamagnetic behavior
for $b \leq 1$. Note, however, that $f(b)$ itself is finite at all $b$, and
thus the Landau diamagnetism prevails.

\subsection{Effect of finite $\mu$}

When the doping is small, $\mu$ is small
as well and corresponds to an
effective quasiparticle description. First, consider $T=0$; as long as
$\mu$ is small enough for the linearization of the spectrum to be valid, the
Fermi surface changes from a point in momentum space for $\mu=0$ to a circle,
and generates a length scale of $\lambda=\hbar v_F/\mu$, the inter electronic
spacing. In the limit $\lambda > l_B$ we get from Eq.~(\ref{eq:Ommu0})
\begin{equation}
\chi_0(\mu)\sim -\frac{1}{(B+\mu^2/\alpha)^{1/2}}
\end{equation}
in the leading order. It is now
obvious that for $\lambda \gg l_B$ we get  $\chi
\sim B^{-1/2}$. This divergence of $\chi$ 
is cut off for $l_B \geq \lambda$,~\cite{Nersesyan,Nguyen,Cangemi,Gusynin} and we get
\begin{equation}
\chi_0(\mu,B = 0)=-\frac{C\alpha}{12\mu}.
\end{equation}
This is of course expected because the
chemical potential tunes the system away from the quantum criticality. For finite
$B$, in the (non-critical) regime of $\lambda < l_B$ we expect de-Haas Van
Alphen (dHVA) oscillation in the magnetization~\cite{Nguyen,Cangemi,Gusynin}
due to the  cutoff introduced by $\mu$.

For $T\neq0$, the additional $T$-dependent part in Eq.~(\ref{eq:OmmuT})
becomes important; see Eq.~(\ref{eq:OmmuT2}). Because we now
have three different length scales: $\lambda_T$, $l_B$ and $\lambda$, the
expression for $\Omega$ (and $\chi$) will depend on their relative magnitudes.
The most important regime from the perspective of criticality, $l_B \ll \lambda
\ll \lambda_T$, is particularly simple. In this case we can use similar
approximations as in Eq.~(\ref{eq:XTlgb}), yielding:
\begin{eqnarray}
&&\chi(l_B \ll \lambda \ll \lambda_T)=\chi_0\bigg[1+\frac{1}{b}
\bigg\{\cosh(\lambda_T/\lambda) \nonumber\\
&\times&\left(F_{3/2}(b)-b^2 F_{5/2}(b)+\frac{b^4}{12} F_{7/2}(b)\right)\bigg\}\bigg].
\label{eq:Xmu2}
\end{eqnarray}
Thus, the susceptibility has a scaling form in terms of two independent
dimensionless variables: $\lambda_T/\lambda$ and $b=\lambda_T/l_B$. The  expression for $\chi$ in Eq.~(\ref{eq:Xmu2}) is valid even if $l_B \ll
\lambda_T < \lambda$, but the latter condition invalidates the applicability
of the linearized theory due to large $\mu$.

In the opposite limit of linear response, simple expressions for the
susceptibility can be derived, and we get
\begin{equation}
\chi(\lambda_T \ll l_B,\lambda)=-N_{f}\frac{\mathrm{\rho_{2D}}\mu_B^2}{3k_{B}T}{\rm sech}^2\left(\frac{\lambda_T}{2\lambda}\right)
\label{eq:Xmu3}
\end{equation}
which reducuces to Eq.~(\ref{eq:Xsmallb}) when $\mu=0$, as expected.

\section{Three dimensions: effect of weak interlayer coupling}
Materials where this $2D$ nodal fermion theory is applicable are layered (quasi-$2D$) systems, an exception being graphene, which is indeed atomically thin.  If we include weak interlayer coupling in a tight binding hamiltonian, the energy spectrum acquires an additional  quadratic dispersion given by 
\begin{equation}
E(\vec{k})={t_{\perp}}k_{z}^2\ell^{2}\pm\hbar v_{F}\sqrt{k_{x}^2+k_{y}^{2}}
\end{equation}
 where $t_{\perp}$ is the interlayer hopping matrix element and $\ell$ is the interlayer spacing. Introduction of this new energy scale will cut the divergence off  $\chi(T\rightarrow0)$  when the magnetic energy scale becomes smaller than $t_{\perp}$.   The corresponding Landau energy spectrum is
 \begin{equation}
E_{n}(k_{z})={ t_{\perp}}k_{z}^2\ell^{2} \pm \sqrt{\alpha Bn}
\end{equation}
For $T=0$, $\mu=0$ limit, we get
\begin{equation}
\frac{2\pi \Omega_{0}^{3D}}{CB}=\int _{-\pi/\ell}^{\pi/\ell} dk_{z} \sum_{n=\tilde{n}_{c}+1}^{\infty}\left[ t_{\perp}k_{z}^2\ell^{2} - \sqrt{\alpha Bn}\right]
\end{equation}
where $\tilde{n}_{c}=\mathrm{Int}\left[({t_{\perp}}k_{z}^2\ell^{2}/\sqrt{\alpha B})^2\right]$. Performing the $n$ sum we get
\begin{equation}\begin{split}
\Omega_{0}^{3D}&=\frac{CB}{2\pi}\int_{-\pi/\ell}^{\pi/\ell} dk_{z}\bigg[{t_{\perp}}k_{z}^2\ell^{2}\zeta(0, \tilde{n}_{c}+1)\\
&-\sqrt{\alpha B}\zeta(-1/2,\tilde{n}_{c}+1)\bigg]
\end{split}
\end{equation}
If $(t_{\perp}k_{z}^2\ell^{2}/\sqrt{\alpha B})^2<1$ for any value of $k_{z}$ within the cutoff, $\tilde{n}_{c}=0$, and the $k_{z}$ integrals can be done trivially. Thus,
\begin{equation}
\Omega_{0}^{3D}=-\frac{CB\pi^{2}{t_{\perp}}}{6\ell}-\frac{C\sqrt{\alpha}B^{3/2}\zeta(-1/2)}{\ell}
\label{eq:3D}
\end{equation}
The susceptibility now is just the previous zero temperature result divided by $\ell$. This implies 
$\hbar v_{F}/{l_{B}t_{\perp}}<\pi^2$, which leads to a lower  cutoff in  the magnetic field, given by $B_{c}=\pi^{4}{ t_{\perp}^2}c/(2e\hbar v_{F}^2)$. For a given $t_{\perp}$, and  $B>B_{c}$,  $\chi \sim B^{-1/2}$. When $ t_{\perp}$ is vanishingly small,  $B_{c}$ is also vanishingly small and can be ignored.
 When $(t_{\perp}k_{z}^2\ell^{2}/\sqrt{\alpha B})^2>1$ for any value of $k_{z}$, the result  is more complicated and will be representative of  a truly 3D system.

However, in $3D$ electrodynamics one has to distinguish between $B$ and $H$, which leads to another cutoff. Following Ref.~\onlinecite{Nersesyan},  we provide the appropriate formulas for $\chi$ at $\mu=0$. 
In $3D$ electrodynamics the magnetic induction $B$ and $H$ must be distinguished:
\begin{equation}
B=H+4\pi M_{3D}(B) .
\end{equation}
For  $\chi$ we must find $B$ as a function of $H$. Since, $M$ in general is a function of $B$ and $T$,  $B$ is a function of $H$ and $T$.  From Eq.~(\ref{eq:om0}) we get for $\lambda_{T}\gg l_{B}$
\begin{equation}
M_{3D}(T=0)=-2{\cal N}_{0}N_{f}
g_{\mathrm{2D}}\mu_{B}^{2}\sqrt{B_0B}/\ell
\label{eq:M3D}
\end{equation}
 Now using Eq.~(\ref{eq:M3D}) we obtain the relation between $B$ and $H$:
\begin{equation}
B(H,T=0)=[(H+H_*)^{1/2}-H_*^{1/2}]^2
\label{eq:BandH}
\end{equation}
where,
\begin{equation}
H_*=(4\pi{\cal N}_{0}N_{f}g_{\mathrm{2D}}\mu_{B}^{2}\sqrt{B_0}/\ell)^2
\label{eq:Hstar}
\end{equation}
and has the dimension of $H$. Plugging
Eq.~(\ref{eq:BandH}) into Eq.~(\ref{eq:M3D}) we get,
\begin{equation}
\chi_{3D}(T=0)=\frac{\partial M_{3D}}{\partial H}=-\frac{1}{4\pi}\left(1+\frac{H}{H_*}\right)^{-1/2}
\label{eq;X3D}
\end{equation}
The same analysis in the linear response regime, $\lambda_{T}\ll l_{B}$, yields
\begin{equation}
B(H,\lambda_{T}\ll l_{B})=\frac{ H}{1+T_0/T}
\end{equation}
and
\begin{equation}
\chi_{3D}(\lambda_{T}\ll l_{B})=-\frac{1}{4\pi}\frac{1}{1+T/T_0}
\label{eq:linX3D}
\end{equation}
where $T_0=4\pi N_{f}\rho_{\mathrm{2D}}\mu_B^2/3k_B\ell$.

\section{Experimental Relevance}

We have established in Sec.~\ref{sec:result} that the diamagnetic susceptibility undergoes
a crossover as a function of $T$, from its zero temperature power-law behavior ($\chi \sim -B^{-1/2}$)
to high temperature
linear behavior ($\chi \sim -1/T$). It is interesting to ask if this crossover  is
observable. Consider graphene; we take the
experimental value of $v_F=10^8$ cm/s and use Eq.~(\ref{eq:Xsmallb}) for the high-$T$ regime. We obtain 
\begin{equation}
\chi_{\mathrm{2D}}=-\frac{9.88\times 10^{-10}}{T} ~{\mathrm{emu/cm}}^2, 
\end{equation}
where the temperature $T$ must be expressed  in Kelvin. Note that $\chi_{\mathrm{2D}}\equiv \chi$ of  Eq.~(\ref{eq:netchi}).

In order to compare with experiments on layered (quasi 2D) materials we calculate  the susceptibility by dividing  Eq.~\ref{eq:Xsmallb} by $\ell$.  If we now take $\ell=3.35 \; {\mathrm{\AA}}$ (the value for graphite), we obtain 
\begin{equation}
\frac{\chi_{\mathrm{2D}}}{\ell} \approx -\frac{2.95\times 10^{-2}}{T}. 
\end{equation}
The corresponding  susceptibility per unit mass is
\begin{equation}
\frac{\chi_{\mathrm{2D}}}{\ell \rho_{3D}^m}=-\frac{0.0134}{T}\mathrm{emu/gm},
\end{equation} 
using the mass density of graphite\cite{McClure} $\rho_{3D}^m=2.22$ ${\mathrm{g/cm^3}}$, which agrees very well with the experimental results.
\cite{Ganguli,McClure}
\begin{figure}[htb]
\includegraphics[scale=0.425]{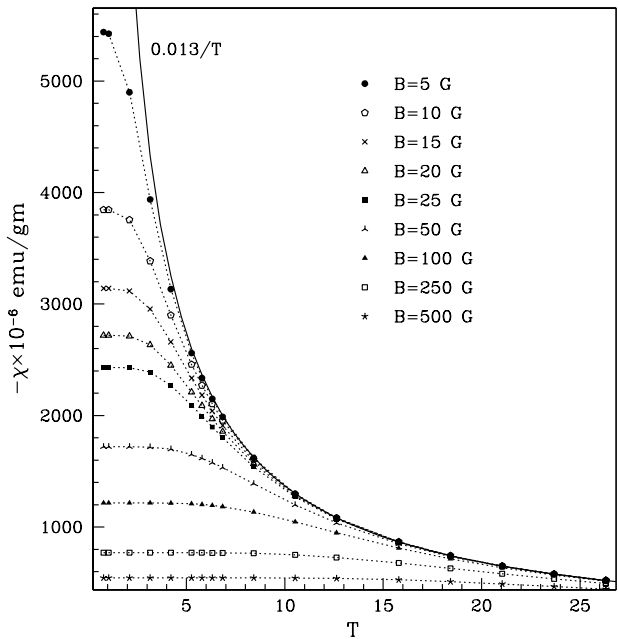}
\caption{
Evolution of $\chi\equiv\chi_{\mathrm{2D}}/\ell \rho_{3D}^m$ (defined in text) as a function of $T$ for various values of  $B$. It is calculated for graphene, using $v_F=10^8$ cm/s, $\ell=3.35{\mathrm{\AA}}$ and $\rho_{3D}^m=2.22{\mathrm{gm/cm^3}}$ }
\label{fig:Fig2}
\end{figure}

Upon lowering $T$,  $\chi_{\mathrm{2D}}$ is strongly enhanced and the power-law region can be accessed for
$\lambda_{T}>l_{B}$. This implies that in graphene for $B>5.6\times 10^{-2}T^{2}$~(Gauss) we must use
Eq.~(\ref{eq:XTlgb}) instead of Eq.~(\ref{eq:Xsmallb}) for the estimation of $\chi_{\mathrm{2D}}$. In particular,
in the $T\rightarrow 0$ limit we obtain 
\begin{equation}
\frac{\chi_{\mathrm{2D}}}{\ell \rho_{3D}^m}=-\frac{0.012}{B^{1/2}}~\mathrm{emu/gm}.
\end{equation}
We demonstrate in Fig. 2 the behavior of $\chi_{\mathrm{2D}}/\ell \rho_{3D}^m$ as a function of $T$ by
numerically evaluating Eq.~(\ref{eq:netchi}) to illustrate the aforementioned crossover behavior.

However, if we use Eq.~(\ref{eq;X3D}) to take into account the demagnetization effect due to interlayer coupling in 3D graphite, we obtain 
\begin{equation}
\chi_{\rm 3D}\approx -\frac{2.95\times 10^{-2}}{(T+0.37)}
\end{equation} 
and this is a very small effect for high temperatures. For 3D graphite, when $\lambda_{T}>l_{B}$, we use
 Eq.~(\ref {eq:Hstar}) and Eq.~(\ref{eq;X3D}). For graphite $H_*=0.11$~G and 
 \begin{equation}
 \chi_{\rm 3D}=-\frac{1}{4\pi}\frac{1}{(1+9.1H)^{1/2}}.
 \end{equation} 
 Therefore, in the limit of $H\ll 0.11$~Gauss  graphite should  become a perfect diamagnet (!), which, however, is a very small field. If the condition, $\lambda_{T}> l_{B}$ is combined with the value of the scale $H_*$, we find that when $T\leq 1.5$K, the demagnetization effect will be important.

In Sec.~\ref{subsec:ddw} we described  the DDW phase of high $T_c$
superconductors by the Hartree-Fock theory of the nodal fermions in the 
copper-oxide layers. We shall estimate the strength of the
diamagnetic susceptibility from the DDW order, using the following experimental
parameters for typical cuprates~\cite{Zhou, Ong} $v_F=3\times10^7$ cm/s and $\ell=12 {\mathrm{\AA}}$, where
$v_{\mathrm{DDW}}$ is estimated assuming a fully formed DDW gap $W_0 \approx 35$ meV
which leads to the anisotropy in the velocity $v_F/v_{\rm DDW}\approx28.6$
if $t=250$ meV.

In the linear response ($\chi \sim -1/T$) regime, we obtain from Eq.~(\ref{eq;X3D}), 
\begin{equation}
\chi_{3D}\approx-\frac{2.6 \times 10^{-5}}{(T+3.3\times 10^{-4})} \approx -\frac{2.6 \times 10^{-5}}{T}.
\end{equation}
When $\lambda_{T}>l_{B}$, we use Eq.~(\ref {eq:Hstar}) and Eq.~(\ref{eq;X3D}) to obtain $H_*=2.7\times 10^{-5}$~G and 
\begin{equation}
\chi_{\rm 3D}=-\frac{1}{4\pi}\frac{1}{(1+3.7\times 10^4H)^{1/2}}.
\end{equation}
This indicates that the diamagnetic susceptibility of DDW from nodal fermions may be measurable. The discussion above would  imply that  DDW would become a perfect
diamagnet when $H \ll 2.7\times 10^{-5}$~G, which, however, is such a  small field that many other effects will intervene, and one would observe  $\chi \sim -H^{-1/2}$, but not perfect diamagnetism. 

In the experiment on BSCCO \cite{Ong} a $T$-dependent power law is observed over a wide range of temperature in the small $H$ limit. Moreover, at the smallest value of the magnetic field in Ref.~\onlinecite{Ong}, $H=5$~G, we get from the DDW calculations  $\chi_{3D}\approx-1.9\times 10^{-4}$ in CGS units as $T\rightarrow 0$.  This is orders of magnitude smaller than that found in the experiment. Therefore, the magnitude of the diamagnetic susceptibility of Ref.~\onlinecite{Ong} can not be explained within a DDW framework alone. One must note however that as the temperature is lowered, the system will generically enter from the DDW phase to a coexisting DDW and $d$-wave superconducting phase for much of the parameter regime.~\cite{Chakravarty} Thus it is clear that the superconducting diamagnetic effects of the Kosterlitz-Thouless theory cannot be ignored.~\cite{Sondhi} But of course none of these considerations can explain the observed critical phase, which requires new ideas.

\section{Conclusion}

We have shown that the notion of quantum criticality, although restricted to non-interacting nodal fermions as elementary excitations, offers interesting insights to diamagnetism of semimetals. When the the chemical potential is zero, the system is inherently
quantum critical, and we derived the scaling function for $\chi$. The
scaling form suggests that the non-linear behavior of $\chi$ as a
function of $B$, due to quantum criticality, can persist up to a large enough temperature, which may be accessible in measurements in graphene.  We have also discussed  how $\mu$ tunes the system away from the quantum critical region. The root of the large magnitude of the diamagnetic susceptibility in graphene or graphite is of course the large Fermi velocity $v_{F}$.  

There are a number of difficult but obvious questions regarding the roles of electron-electron interaction and disorder. These could be topics for future work. We have seen that our simple picture of the DDW does not explain the remarkable experiments in the high temperature superconductors. We do not know if the generalization of the Hartree-Fock picture of the DDW to the six-vertex model where a power law high temperature phase was found~\cite{sixvertex} will be able to explain these experiments. It is certainly worth exploring.  We stress, for the reasons stated above, that these experiments are not fully explained by Kosterlitz-Thouless theory, as is sometimes claimed. 
 
It is clear that the Euler-MacLaurin summation approach to compute Landau diamagnetism for non-relativistic fermions fails because of the non-analyticity due to massless Dirac fermions  in semimetals. It is not known to us if there are any systems for which $(3+1)$-dimensional quantum critical  behavior $\chi\sim \log H$ is experimentally observable. The material   $\mathrm{Bi_{1-x}Sb_{x}}$  is lamellar, as is bismuth telluride, and is better described as a two-dimensional system with weak interlayer coupling. Nonetheless, it would be interesting to study the diamagnetism of this material as a function composition.

\begin{acknowledgments}
We thank Stuart Brown and Subir Sachdev for discussions. This work was in part carried out at the Aspen Center for Physics and was supported by the NSF under the Grant No. DMR-0411931, and also by the funds from the David Saxon Chair at UCLA.
\end{acknowledgments}

\appendix
\section{High temperature series: $\mu\ne 0$}
\begin{widetext}
The grand canonical thermodynamic potential is given by
\begin{equation}
\Omega(T,\mu)=-CBk_{B}T\left[\log\left\{2\cosh\left(\frac{\mu}{2k_{B}T}\right)\right\}+\sum_{n=1}^{\infty}\log\left\{2\cosh\left(\frac{\mu-E_{n}}{2k_{B}T}\right)\right\}+\sum_{n=1}^{\infty}\log\left\{2\cosh\left(\frac{\mu+E_{n}}{2k_{B}T}\right)\right\}\right].
\label{eq:A1}
 \end{equation}
 \end{widetext}
Each individual LL sum  fails standard convergent tests.
The 
technique to deal with such sums in the quantum critical regime is discussed in the text.
The strategy in the other limit, where linear response holds, is to convert the
LL sums to express them as series in powers of $b \sim 1/T$, so that meaningful
conclusions could be drawn about the small $b$ limit (equivalently, high $T$
limit) by considering leading order terms systematically. The procedure relies
on $\zeta$-function regularization, details of which could be found in
literature, but the purpose of this appendix is to provide a self contained 
description.
Separating out the zero temperature part $\Omega_0(\mu)$ and finite
temperature part $\Omega_T(\mu)$ we obtain Eq.~(\ref{eq:Ommu0}) and
Eq.~(\ref{eq:OmmuT}) respectively. We now wish to express $\Omega_T(\mu)$
as a series expansion in powers of $b$. For this purpose we focus below to
one term in Eq.~(\ref{eq:OmmuT}), say, the following one:
 \begin{equation}\begin{split}
I&=\sum_{n=1}^{n_c} \log(1+e^{-\frac{\mu-E_n}{k_{B}T}}) \\
&=\sum_{n=1}^{\infty} \log(1+e^{-\frac{\mu-E_n}{k_{B}T}})-\sum_{n_c+1}^{\infty} \log(1+e^{-\frac{\mu-E_n}{k_{B}T}}).
\label{eq:A3}
\end{split}
\end{equation}
We will now expand both the summations (we call them $I_1$ and $I_2$
respectively), first the logarithms in powers of the exponentials and
subsequently $e^{-\frac{\mu-E_{n}}{T}}$ in a power series to write
\begin{eqnarray}
I_1&=&\sum_{n=1}^{\infty} \log(1+e^{-\frac{\mu-E_n}{k_{B}T}}) \nonumber\\
&=&-\sum_{n=1}^{\infty} \sum_{k=1}^{\infty} \frac{(-e^{-\frac{\mu}{k_{B}T}})^k}{k} \sum_{r=0}^{\infty} \frac{b^r}{r!} k^r n^{\frac{r}{2}} \nonumber\\
&=&-\sum_{r=0}^{\infty} \frac{b^r}{r!} \sum_{k=1}^{\infty} \frac{(-e^{-\frac{\mu}{k_{B}T}})^k}{k^{1-r}} \sum_{n=1}^{\infty} n^{\frac{r}{2}} \nonumber\\
&=&\sum_{r=0}^{\infty} \frac{b^r}{r!} {\rm Li}_{1-r}(-e^{-\frac{\mu}{k_{B}T}}) \zeta(-r/2) \cdot
\label{eq:A4}
\end{eqnarray}
In the third step above, we interchanged the order of summation, which in
general leads to a correction, but in this particular case it
 is zero (for details see Ref.~ \onlinecite{Elizalde}). And in the final
step we have used the standard definition of Polylogarithm
${\rm Li}_s(z)=\sum_{k=0}^{\infty} z^k/k^s$ and the Riemann
$\zeta$-function. Similar manipulations for $I_2$ lead to the following:
\begin{widetext}
\begin{equation}
I_2=\sum_{n_c+1}^{\infty} \log(1+e^{-\frac{\mu-E_n}{k_{B}T}}) 
=-\sum_{r=0}^{\infty} \frac{b^r}{r!} \sum_{k=1}^{\infty} \frac{(-e^{-\frac{\mu}{k_{B}T}})^k}{k^{1-r}} \sum_{n_c+1}^{\infty} n^{\frac{r}{2}}
=\sum_{r=0}^{\infty} \frac{b^r}{r!} {\rm Li}_{1-r}(-e^{-\frac{\mu}{k_{B}T}}) \zeta(-r/2, 1+n_c)
\label{eq:A5}
\end{equation}
where we get Hurwitz's $\zeta$-function instead of Riemann $\zeta$-function.
Employing similar simplification to each term of Eq.~(\ref{eq:OmmuT}), we
finally obtain the desired high temperature series expansion:
\begin{eqnarray}
\Omega_T(\mu)&=&-CBk_{B}T\bigg[
\log\left(1+e^{-\frac{\mu}{k_{B}T}}\right)-\sum_{r=0}^{\infty}\frac{b^r}{r!}{\rm Li}_{1-r}(-e^{-\frac{\mu}{k_{B}T}})
[1+(-1)^r]\zeta(-\frac{r}{2})
+\sum_{r=0}^{\infty}\frac{b^r}{r!}\zeta(-\frac{r}{2},1+n_c)  \nonumber \\
&& \times  \left\{ {\rm Li}_{1-r}(-e^{-\frac{\mu}{k_{B}T}})+
(-1)^r{\rm Li}_{1-r}(-e^{\frac{\mu}{k_{B}T}})\right\}
\bigg]
\label{eq:OmmuT2}
\end{eqnarray}
\end{widetext}
We arrive at Eq.~(\ref{eq:fsum})
by letting $\mu=0$.  Also, starting from Eq.~(\ref{eq:OmmuT2}) we can
derive Eq.~(\ref{eq:Xmu3}).

\section{Sum for $\lambda_T \gg l_B$}
When $b \gg 1$, one can simplify in Eq.~(\ref{eq:Ssum}) $S=\partial I/\partial b$, where
\begin{equation}
I=\sum_{n=1}^{\infty} e^{-b\sqrt{n}}=\sum_{n=1}^{\infty} \sum_{r=0}^{\infty}
\frac{(-b)^r}{r!}n^{r/2}.
\label{eq:Idef}
\end{equation}
Using the integral representation of the $\Gamma$-function in
Eq.~(\ref{eq:Idef}) we have
\begin{equation}
I=\sum_{r=0}^{\infty} \frac{(-b)^r}{r!}\frac{1}{\Gamma(-r/2)}\int_{0}^{\infty}
dx x^{-r/2-1} \sum_{n=1}^{\infty}e^{-nx} \cdot
\label{eq:step2}
\end{equation}
Note that the change  of the order of sum and integral does not
result in any extra terms.~\cite{Elizalde} The sum over $n$ can now be trivially
performed and using the relation
\begin{equation}
\frac{1}{\Gamma(-r/2)}=-\frac{\Gamma(1+r/2)}{\pi}\sin(\frac{\pi r}{2}),
\label{eq:step3}
\end{equation}
we get
\begin{equation}\begin{split}
I&=-\frac{1}{\pi} \int_0^{\infty} \frac{dx}{x(e^x-1)} \sum_{r=1}^{\infty}
\frac{(-1)^r}{\Gamma(r+1)} \left(\frac{b}{\sqrt{x}}\right)^r \\
\times& \sin\left(\frac{\pi r}{2}\right) \Gamma \left(1+\frac{r}{2}\right) .
\label{eq:step4}
\end{split}
\end{equation}
After carrying out the $r$-sum we get
\begin{equation}
I=\int_0^{\infty} dx \frac{e^{-b^2/4x} x^{-3/2}}{e^x-1} .
\label{eq:step5}
\end{equation}
This gives  $\chi_T$ in Eq.~(\ref{eq:XTlgb}). Alternatively, we could have
expanded the logarithm term in $\Omega_T$ and have kept only the first term in
that expansion for $b \gg 1$; one has $\Omega_T=-CTBI$. Thus, one arrives at
the same expression for $\chi_T$ as in Eq.~(\ref{eq:XTlgb}).

For the saddle-point approximation of $F_p(b)$ as defined in Eq.~(\ref{eq:Fdef}),
we write:
\begin{equation}\begin{split}
F_p&=\int_0^{\infty}dx g(x) e^{-h(x)} \\
&\approx g(x_0) e^{-h(x_0)} \int_0^{\infty}dx \exp\left[-\frac{h''(x_0)}{2} (x-x_0)^2\right]
\label{eq:A6}
\end{split}
\end{equation}
with $g(x)=[\sqrt{\pi}(e^x-1)]^{-1}$ and $h(x)=b^2/4x+p\log x$, and $x_0$ is
defined by $h'(x_0)=0$ (the prime refers to derivative). Simple manipulations following this
scheme yield Eq.~(\ref{eq:XTlgb}) and Eq.~(\ref{eq:Ndef}).

\end{document}